\documentclass[prl,twocolumn,showpacs,preprintnumbers,amsmath,amssymb]{revtex4}
\usepackage{graphicx}
\usepackage{dcolumn}
\usepackage{epsfig}

\begin{document}
\title{High field Hall resistivity and magnetoresistance in electron-doped Pr$_{2-x}$Ce$_{x}$CuO$_{4-\delta}$}
\author{Pengcheng Li$^1$}
\author{F. F. Balakirev$^2$}
\author{R. L. Greene$^1$}
\affiliation{$^1$Center for Superconductivity Research and
Department of Physics, University of Maryland, College Park,
Maryland 20742-4111 \\$^2$NHMFL, Los Alamos National Laboratory,
Los Alamos, NM 87545}\

\date{\today}

\begin{abstract}
We report resistivity and Hall effect measurements in
electron-doped Pr$_{2-x}$Ce$_{x}$CuO$_{4-\delta}$ films in
magnetic field up to 58 T. In contrast to hole-doped cuprates, we
find a surprising non-linear magnetic field dependence of Hall
resistivity at high field in the optimally doped and overdoped
films. We also observe a crossover from quadratic to linear field
dependence of the positive magnetoresistance in the overdoped
films. A spin density wave induced Fermi surface reconstruction
model can be used to qualitatively explain both the Hall effect
and magnetoresistance.
\end{abstract}
 \maketitle
Electron-doped (n-doped) cuprate superconductors have exhibited
enough similarities with their hole-doped (p-doped) high-T$_c$
counterparts so that any eventual rationalization of the
phenomenon of high temperature superconductivity (SC) would have
to treat both poles of the doping spectrum in the similar manner.
Some of the key phenomena realized in both types of high-T$_c$
compounds, such as the competition between antiferromagnetism(AFM)
and SC or the anomalous temperature dependence of the transport
coefficients, pose challenging questions for the condensed matter
physics. Meanwhile, a number of studies in the past years have
identified distinct differences in the properties of n-doped and
p-doped cuprates. Understanding the causes of the differences and
similarities may lead to understanding of the phenomenon of high
temperature superconductivity. Among the distinctive properties,
angle resolved photoemission spectroscopy (ARPES)
experiments~\cite{Armitage,Matsui,Matsui2} in n-doped cuprates
have revealed a small electron-like Fermi surface (FS) pocket at
$(\pi, 0)$ in the underdoped region, and a simultaneous presence
of both electron- and hole-like pockets near optimal doping. This
clarifies the long-standing puzzle that transport in these
materials exhibits unambiguous n-type carrier behavior at
low-doping, and two-carrier behavior near optimal
doping~\cite{Ong,Jiang,Fournier,Gollink}. Recent low temperature
normal state Hall effect measurements~\cite{Yoram} on
Pr$_{2-x}$Ce$_{x}$CuO$_{4-\delta}$ (PCCO) show a sharp kink of the
Hall coefficient at a critical doping \textit{x}=0.16, which
suggests a quantum phase transition (QPT) at this doping. Early
$\mu$SR measurements found that the AFM phase starts at
\textit{x}=0 and persists up to, or into, the SC dome~\cite{Luke}.
Neutron scattering experiments~\cite{Kang} show an AFM phase above
critical field in an optimally doped n-type cuprates, but no such
phase on the overdoped side. Optical conductivity
experiments~\cite{Zimmers} reveal a partial normal state gap
opening at a certain temperature in the underdoped region, but no
such gap is found above the critical doping. A spin density wave
(SDW) model~\cite{Lin, Millis} was proposed, which gives a
plausible, but qualitative, explanation to these observations. In
this model, SDW ordering would induce a Fermi surface (FS)
reconstruction and result in an evolution from an electron pocket
to the coexistence of electron- and hole-like pockets with
increasing doping, and eventually into a single hole-like FS. The
SDW gap amplitude decreases from the underdoped side and vanishes
at the critical doping.

Recently, an inelastic neutron scattering measurement on
Nd$_{2-x}$Ce$_x$CuO$_{4-\delta}$ (NCCO) by Motoyama \textit{et
al.}~\cite{Greven} found that long range ordered AFM vanishes near
the SC dome boundary, \textit{x}=0.13. They claim that SC and AFM
do not coexist and that the QPT occurs at \textit{x}=0.13, which
differs from the results mentioned above. Therefore, the presence
of a QPT and the location of the quantum critical point (QCP) in
n-doped cuprates is still under considerable debate. Whether the
QCP is under the SC dome and if SC and AFM coexist are important
issues to clarify. Similar issues exist for the p-doped cuprates,
although the nature of any order parameter (in the pseudogap
phase) which might disappear at a QCP under the SC dome is
unknown.

In this Letter, for the first time, we used a high magnetic field
to investigate the transport behavior in n-doped PCCO over a wide
range of doping and temperature. We performed Hall resistivity and
ab-plane resistivity measurements on PCCO films from underdoped
(UND) (\textit{x}=0.11) to overdoped (OVD) (\textit{x}=0.19) for
temperature down to T=1.5 K and magnetic field up to 58 T. We find
that both the Hall resistivity and magnetoresistance(MR) show a
dramatic change near optimally doped (OPD) \textit{x}=0.15. A
surprising non-linear Hall resistivity for \textit{x}$\geq$0.15 is
observed, while linearity of the Hall resistivity persists up to
58 T in the UND films. In the OVD region, a crossover of MR from a
low field quadratic behavior to a high field linear dependence is
found, which is in agreement with a recent theory on the
magnetotransport properties near a metallic QCP~\cite{Schofield}.
Our results suggest that the QPT occurs around optimal doping in
the n-doped cuprates.

PCCO c-axis oriented films with various Ce concentrations
(\textit{x}=0.11 to 0.19) were fabricated by pulsed laser
deposition on (100) oriented SrTiO$_3$ substrates~\cite{Maiser}.
Since the oxygen content has an influence on both the SC and
normal state properties of the material~\cite{Jiang}, we optimized
the annealing process for each Ce concentration as in
Ref.~\cite{Yoram}. The sharp transition, low residual resistivity
and the Hall coefficient are exactly the same as in our previous
report~\cite{Yoram}. We note that the exact content of oxygen
cannot be determined in films and we use the T$_c$ values of the
annealed films and Ce content to determine the temperature versus
doping phase diagram. Photolithography and ion mill techniques are
used to pattern the films into a standard six-probe Hall bar.
Resistivity and Hall effect measurements were carried out in a 60
T pulsed magnetic field at the NHMFL. The magnetic field is
aligned perpendicular to the ab-plane of the films. Possible eddy
current heating was carefully considered and reduced.
\begin{figure}
\includegraphics[width=9cm, height=4.5cm]{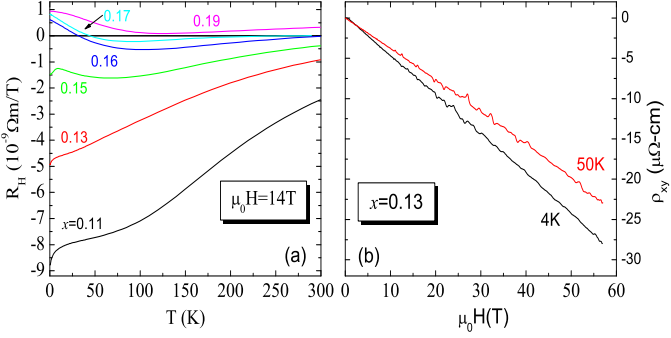}
\caption{(color online). (a) Normal state Hall coefficient $R_H$
versus temperature for PCCO films measured at $\mu_0$H=14 T
($H\perp ab$ and H$>$H$_{c2}$). (b) Hall resistivity $\rho_{xy}$
versus H for the UND PCCO \textit{x}=0.13 film at T=4 K and 50 K.}
\label{Fig1}
\end{figure}

\begin{figure}
\includegraphics[width=9cm, height=8cm]{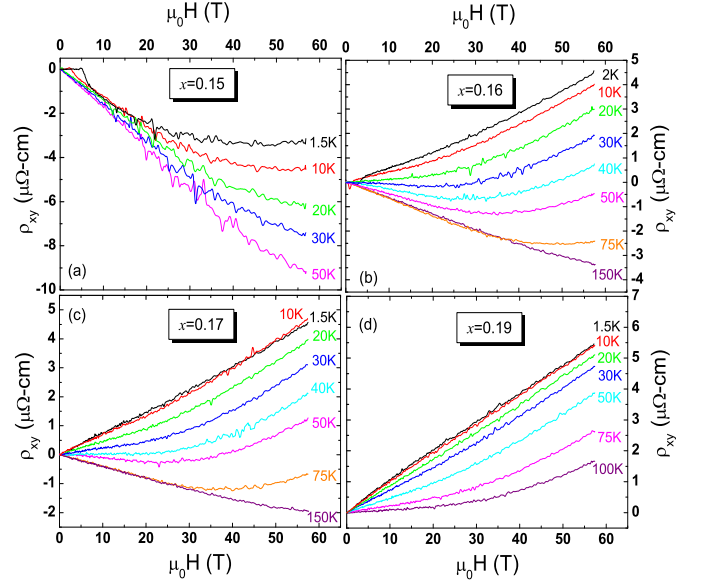}
\caption{(color online). Hall resistivity $\rho_{xy}$ versus H for
the OPD and OVD PCCO films (a) \textit{x}=0.15, (b)
\textit{x}=0.16, (c) \textit{x}=0.17, (d) \textit{x}=0.19.}
\label{Fig2}
\end{figure}

The Hall effect is measured for all the doped PCCO films. At low
field for H up to 14 T, the Hall resistivity $\rho_{xy}(H)$ is
linear at all dopings and temperatures~\cite{Yoram} and the Hall
coefficient $R_H$ is determined as $\rho_{xy}/H$.
Fig.~\ref{Fig1}(a) shows the temperature dependence of $R_H$
measured at 14 T for all the PCCO films, similar to our previous
report~\cite{Yoram}. In high fields, we find that the linear
negative $\rho_{xy}(H)$ (electron-like) persists up to 58 T in the
UND films at temperatures below 100 K (the limit of our present
data). Fig.~\ref{Fig1}(b) shows the $\rho_{xy}(H)$ data for an UND
\textit{x}=0.13 film (T$_c$=11 K). For the non-superconducting
\textit{x}=0.11, a similar behavior is observed (not shown). In
contrast, as \textit{x} approaches the OPD \textit{x}=0.15,
$\rho_{xy}(H)$ behaves differently. As shown in
Fig.~\ref{Fig2}(a), at low temperatures (T$<$50 K) $\rho_{xy}(H)$
is negative and linear up to about 30 T but then starts curving
towards positive slopes. This nonlinearity begins at higher fields
as the temperature is increased and it appears that $\rho_{xy}(H)$
will become linear above 50 K, but we were not able to obtain data
above 50 K on this film. For the OVD film \textit{x}=0.16, the low
temperature $\rho_{xy}(H)$ is positive and slightly non-linear at
high field. The nonlinearity becomes more prominent near the
temperature where $R_H$ changes sign (about 25 K), but then a
negative linearity of $\rho_{xy}(H)$ is found at higher
temperatures. The positive slope of $\rho_{xy}(H)$ at high field
for T$\leq$75 K indicates a hole-like contribution. A similar
behavior was observed for the \textit{x}=0.17 film as shown in
Fig.~\ref{Fig2}(c). For a highly OVD film, \textit{x}=0.19, in
which $R_H$ is always positive, but with a minimum around 120 K
[see Fig.~\ref{Fig1}(a)], a strong field dependence of
$\rho_{xy}(H)$ is observed above 30 K, while $\rho_{xy}(H)$ is
linear at low temperatures. We note that this non-linear
$\rho_{xy}(H)$ in the OPD and OVD films is in striking contrast to
the linear $\rho_{xy}(H)$ found up to 60 T in p-doped
Bi$_{2}$Sr$_{2-x}$La$_{x}$O$_{6+\delta}$ at all dopings and
temperatures~\cite{Fedor}.

\begin{figure}
\includegraphics[width=9cm, height=8cm]{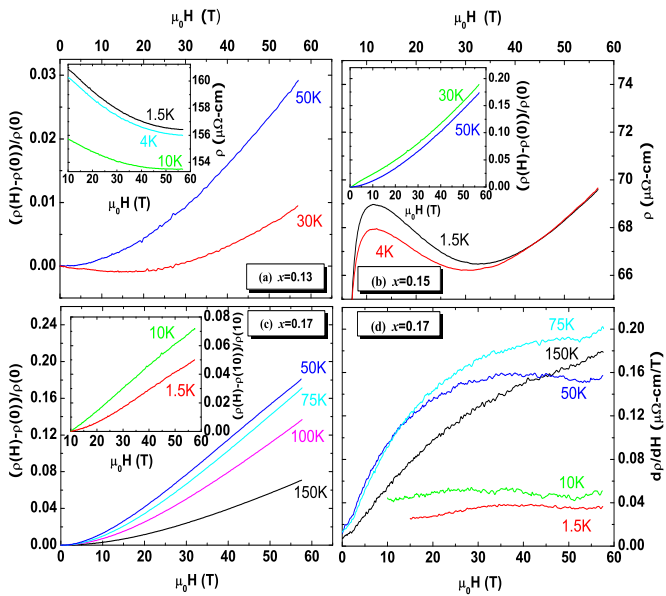}
\caption{(color online). In-plane resistivity versus H ($H\perp
ab$) for PCCO films (a) \textit{x}=0.13 (b) 0.15 (c) 0.17 and (d)
first derivative plots for 0.17. Insets show the magnetoresistance
in a different temperature range from the main panels.}
\label{Fig3}
\end{figure}

The in-plane magnetoresistance (MR) is also measured in PCCO films
and we find that the high field MR varies greatly with doping as
well. In the UND region, a large low temperature negative MR (nMR)
is observed, similar to the prior work~\cite{Yoramupturn}. For
\textit{x}=0.11 (T$<$100 K), this nMR persists up to 58 T, while
for \textit{x}=0.13 (T$<$30 K), the nMR tends to saturate at high
field, as shown in the inset of Fig.~\ref{Fig3}(a). A positive MR
is recovered for T$>$50 K, and it follows a quadratic field
(H$^2$) dependence. For the OPD \textit{x}=0.15, the low
temperature nMR reverses to a positive MR at H$\sim$30 T
[Fig.~\ref{Fig3}(b)]. For T$>$20 K, this positive MR also obeys
$H^2$. In the OVD films, a positive MR is found in the normal
state, but the field dependence of the MR is surprisingly
different as the temperature is increased. Fig.~\ref{Fig3}(c) and
(d) show the MR along with the first field derivative of MR
($d\rho_{xx}(H)/dH$) for \textit{x}=0.17. At low temperatures
(T$<$20 K), a nearly linear MR for H$>$H$_{c2}$ is found, as seen
from the roughly constant behavior of the $d\rho_{xx}(H)/dH$ plot.
However, in the intermediate temperature range where the
non-linear $\rho_{xy}(H)$ is prominent, we find that
$d\rho_{xx}(H)/dH$ increases monotonically at lower field and then
saturates to a nearly constant value at higher field. The low
field linear increase of $d\rho_{xx}(H)/dH$ indicates a quadratic
MR and the high field saturation indicates a linear MR. A similar
linear MR is also observed in the \textit{x}=0.16 and 0.19 films
at low temperatures where the linear $\rho_{xy}(H)$ is in present.
At higher temperatures, the MR changes from quadratic at low field
to linear at high field.

The non-linear behavior of $\rho_{xy}(H)$ displayed on
Fig.~\ref{Fig2} can be simulated within the framework of a
conventional two-band Drude model. The Drude model assumes
field-independent carrier density and relaxation time, in which
case $\rho(H)$ can be written as~\cite{Ashcroft}:
$\rho_{xy}(H)=\frac{\sigma_h^2 R_h-\sigma_e^2
R_e-\sigma_h^2\sigma_e^2 R_h
R_e(R_h-R_e)H^2}{(\sigma_e+\sigma_h)^2+\sigma_e^2\sigma_h^2(R_h-R_e)^2H^2}H$
and $\rho_{xx}(H)$ should be $\rho_{xx}(H)= \frac{(\sigma_h +
\sigma_e)+\sigma_h\sigma_e(\sigma_h R^2_h +\sigma_e
R^2_e)H^2}{(\sigma_h + \sigma_e)^2 + \sigma^2_h \sigma^2_e
(R_h-R_e)^2 H^2}$ [$\sigma_{e(h)}$ and $R_{e(h)}$ are electrical
conductivity and Hall coefficient of the electron (hole) band].
Using the relation of $\sigma_0=\sigma_h+\sigma_e$ ($\sigma_0$ is
the zero field normal state conductivity), one parameter is
eliminated. We attempted to fit our $\rho_{xy}(H)$ and MR data,
but we could not fit both $\rho_{xy}(H)$ and $\rho_{xx}(H)$ with
the same fitting parameters for any of the films. We also find
that the parameters by fitting $\rho_{xy}(H)$ alone are in
conflict with ARPES results. The hole and electron densities at
T=10 K for \textit{x}=0.15 from our fit are
$n_h=\frac{1}{R_h|e|}=6.0\times10^{20}/cm^3$ and $n_e=3.5\times
10^{20}/cm^3$, which disagree with $n_h=3.6\times10^{20}/cm^3$ and
$n_e=1.8\times 10^{21}/cm^3$, the estimate from the areas of the
hole and electron pockets in ARPES~\cite{Armitage, Blumberg}. The
departure of our fits from the experimental data is most likely
due to the simple assumption of a field-independent charge density
and scattering in the Drude model. This is unlikely to be valid
for the n-doped cuprates with their complex FS. It is likely that
there is a strong anisotropic scattering, as found in p-doped
cuprates~\cite{Hussey}. Therefore, the simple Drude model is not
sufficient to explain the high-field magnetotransport in PCCO. A
modified model with consideration of field dependent charge
density or scattering  might explain our data, however, this is
beyond our present knowledge and the scope of this Letter.

\begin{figure}
\includegraphics[width=7cm, height=5.5cm]{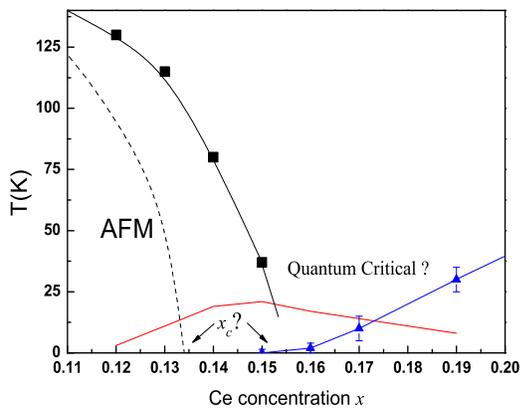}
\caption{(color online). A phase diagram for n-doped cuprates.
Black squares and dash line are the controversial AFM transition
temperature from Ref.~\cite{Luke, Kang} and Ref.~\cite{Greven}
respectively. The red line is T$_c$. The blue triangles mark the
temperature above which the non-linear $\rho_{xy}(H)$ appears for
each doping.} \label{Fig4}
\end{figure}

We now discuss a qualitative (and speculative) explanation for our
data. As seen from the phase diagram of n-doped cuprates (see
Fig.~\ref{Fig4}), the long range ordered AFM phase persists up to
a critical doping of $x_c$ (exact location is under debate). In
the UND region, a large SDW gap ($\triangle_{SDW}$) opens at
certain temperatures~\cite{Zimmers}. In a magnetic field
comparable to the gap (i.e. $\mu_BB\sim \triangle_{SDW}$), one
expects a suppression of $\triangle_{SDW}$ by the field and a
consequent change of the FS. Since $\rho_{xy}(H)$ is sensitive to
the shape of the FS, a non-linear $\rho_{xy}$ in high field might
emerge. Applying this picture to our data, let us start from the
lowest temperature (1.5 K). In the UND \textit{x}=0.11 and 0.13,
we find that the linear $\rho_{xy}(H)$ persists up to 58 T
[Fig.~\ref{Fig1}(b)], suggesting that the field is not sufficient
to destroy the large SDW gap. Therefore, the electron-like pocket
still survives to high field for all the temperatures measured. As
the doping approaches the critical doping, $\triangle_{SDW}$
decreases rapidly. When the magnetic field is strong enough to
suppress the smaller gap, the hole-like pocket emerges and will
contribute to $\rho_{xy}(H)$. The positive slope of $\rho_{xy}(H)$
at high field in the OPD film \textit{x}=0.15 suggests the
suppression of the SDW gap and a contribution from the hole band.

In the OVD region for $x\geq 0.16$, the linear positive
$\rho_{xy}(H)$ at the lowest temperature strongly suggests the
absence of the SDW gap and a hole-like behavior. The nonlinearity
of $\rho_{xy}(H)$ appears at higher temperatures for larger
\textit{x}, as seen in Fig.~\ref{Fig2}. Notice that a slightly
non-linear $\rho_{xy}$(H) is found for \textit{x}=0.16 even at the
lowest temperature 2 K, while in \textit{x}=0.17 and 0.19, this
nonlinearity starts to appear at temperatures above 10 K and 30 K,
respectively, as indicated with blue triangles in Fig.~\ref{Fig4}.

The observed nonlinearity in $\rho_{xy}$(H) in the OVD region in
the intermediate temperature range suggests a competition between
electron and hole bands. This unusual nonlinearity of
$\rho_{xy}$(H) might arise from spin fluctuations of SDW order in
a quantum critical region associated with the QPT at $x_c$. As
shown in Fig.~\ref{Fig4}, thermally activated spin fluctuations
(gap-like) in the OVD region at finite temperature could result in
a mix of electron and hole contributions to $\rho_{xy}(H)$. This
could be responsible for the sign change of the $R_H$(T)
[Fig.~\ref{Fig1}(a)] and the positive upturn of the $\rho_{xy}(H)$
at high fields. In the critical region, an external perturbation,
such as temperature or magnetic field, could change the relative
impact of the two bands. The onset temperature of the high field
non-linear $\rho_{xy}$(H) shifts towards a higher temperature as
\textit{x} increases, which strongly suggests the system is
further away from the critical region at higher doping.

Interestingly, we notice that our MR is also qualitatively
consistent with the SDW model. In the UND region, the low
temperature nMR persists and saturates in high field, suggesting
that the nMR is related to spin scattering~\cite{Yoramupturn}. At
optimal doping, a positive MR is recovered at a field where
$\rho_{xy}(H)$ changes slope. The complete suppression of the nMR
suggests the reduction of the spin contribution. For the OVD films
($x\geq$0.16), we find that the crossover in the field dependence
of the MR is consistent with a recent theory. It has been
shown~\cite{Schofield} that near a metallic QCP, a quadratic MR
($\frac{\triangle\rho}{\rho}\sim H^2$) is expected for a system
with a SDW gap and a linear MR without a gap. As shown in
Fig~\ref{Fig3}(c) and (d), in the intermediate temperature range,
the MR changes from quadratic to linear as the field increases
while it is always linear at low temperature. This suggests the
absence of the SDW gap at low temperature. In the spin fluctuation
region, the field suppressed spin contribution could be
responsible for the recovery of the linear MR at high field. At
much higher temperatures (above 150 K), the quadratic MR is
restored.

We have shown that our $\rho_{xy}(H)$ and $\rho_{xx}(H)$ are
qualitatively consistent with the SDW model, suggesting that the
QCP locates near or just above the optimal doping. However, the
SDW gap closing field (about 30-40 T for \textit{x}=0.15) obtained
from the nonlinear $\rho_{xy}(H)$ is much smaller than that from
the optics measurements~\cite{Zimmers}. Further quantitative
theoretical calculations are needed to resolve this issue and to
explain in detail the $\rho_{xy}(T,H)$ and $\rho_{xx}(T,H)$
results that we report here.

In summary, we performed high field resistivity and Hall effect
measurements in the n-doped cuprate
Pr$_{2-x}$Ce$_{x}$CuO$_{4-\delta}$. We find an anomalous
non-linear Hall resistivity at high field above optimal doping at
certain temperatures. We also observed a crossover of the field
dependence of magnetoresistance at high field in the overdoped
region. Our results are qualitatively consistent with the spin
density wave gap induced Fermi surface rearrangement
model~\cite{Lin}, and also support the view that a quantum phase
transition occurs under the superconducting dome in this material.

We are grateful to the fruitful discussions with A. Millis, V.
Yakovenko and A. Chubukov. P.L. and R.L.G. acknowledge the support
of NSF Grant DMR-0352735. The work at NHMFL was supported by NSF
and DOE.


\begin{references}
\bibitem{Armitage}N. P. Armitage, \textit{et al.},Phys. Rev. Lett. \textbf{88}, 257001
(2002).
\bibitem{Matsui}H. Matsui, \textit{et al.},Phys. Rev. Lett. \textbf{94}, 047005
(2005).
\bibitem{Matsui2}H. Matsui, \textit{et al.},Phys. Rev. Lett. \textbf{95}, 017003
(2005).
\bibitem{Ong}Z. Z. Wang, \textit{et al.}, Phys. Rev. B \textbf{43},
3020 (1991).
\bibitem{Jiang}W. Jiang \emph{et al}., Phys. Rev. Lett.\textbf{73} 1291
(1994).
\bibitem{Fournier}P. Fournier \emph{et al}., Phys. Rev. B \textbf{56}, 14149
(1997).
\bibitem{Gollink}F. Gollnik and M. Naito, Phys. Rev. B \textbf{58}
11734 (1998).
\bibitem{Yoram}Y. Dagan \textit{et al.}, Phys. Rev. Lett. \textbf{92}, 167001
(2004).
\bibitem{Luke}G. M. Luke \textit{et al.} Phys. Rev. B \textbf{42}, 7981
(1990). T. Uefuji \textit{et al.}, Physica \textbf{C378}, 273,
(2002).
\bibitem{Kang}H. Kang \textit{et al.}, Nature(London) \textbf{423}, 522
(2003); M. Fujita \textit{et al.}, Phys. Rev. Lett.
\textbf{93},147003 (2004).
\bibitem{Zimmers} A. Zimmers \textit{et al.}, Europhys. Lett. \textbf{70}, 225
(2005); Y. Onose \textit{et al.}, Phys. Rev. Lett. \textbf{87},
217001 (2001).
\bibitem{Lin}J. Lin and A. J. Millis, Phys. Rev. B \textbf{72}, 214506
(2005).
\bibitem{Millis}A. J. Millis, \textit{et al.}, Phys. Rev. B \textbf{72}, 224517
(2005).
\bibitem{Greven}E. M. Motoyama \emph{et al}., Nature(London) \textbf{445},
186 (2007).
\bibitem{Schofield}J. Fenton and A. J. Schofield, Phys. Rev. Lett. \textbf{95}, 247201
(2005).
\bibitem{Maiser} E. Maiser \textit{et al.}, Physica(Amsterdam) \textbf{297C}, 15
(1998); J. L. Peng \textit{et al.}, Phys. Rev. B \textbf{55} R6145
(1997).
\bibitem{Fedor}F. Balakirev \textit{et al.}, Nature(London) \textbf{424},
912 (2003).
\bibitem{Ashcroft}N. W. Ashcroft and N. D. Mermin, \emph{Solid State Physics} (Saunders College Publishing, 1976),
p.240.
\bibitem{Blumberg}The electron and hole densities are obtained from the size of the electron and hole FS
pockets, i.e. $n=A/2\pi^2$(A is the the area of the electron(hole)
pocket from the ARPES in Ref.~\cite{Armitage} and a SDW model
calculation~\cite{Lin, Millis}), with G. Blumberg and M. Qazilbash
(private communication).
\bibitem{Hussey}M. Abdel-Jawad \textit{et al.}, Nature Physics, \textbf{2},
821 (2006).
\bibitem{Yoramupturn} Y. Dagan \emph{et al}., Phys. Rev. Lett. \textbf{94}, 057005
(2005).
\end{references}
\end{document}